\documentclass[preprint]{revtex4}
\usepackage[latin1]{inputenc}
\usepackage{multirow}
\usepackage{color}
\usepackage[english]{babel}
\usepackage{subfigure}
\usepackage{graphicx}
\usepackage{amsmath}
\usepackage{amsfonts}
\usepackage{amssymb}
\usepackage[ps2pdf]{hyperref}
\hypersetup{colorlinks=true,linkcolor=blue,citecolor=red,pdfpagemode=None}
\newcommand{\promille}{%
  \relax\ifmmode\promillezeichen
        \else\leavevmode\(\mathsurround=0pt\promillezeichen\)\fi}
\newcommand{\promillezeichen}{%
  \kern-.05em%
  \raise.5ex\hbox{\the\scriptfont0 0}%
  \kern-.15em/\kern-.15em%
  \lower.25ex\hbox{\the\scriptfont0 00}}
\def\gsim{\mathrel{\rlap{\lower4pt\hbox{\hskip1pt$\sim$}} \raise1pt\hbox{$>$}}} % greater than or approx. symbol
\def\lsim{\mathrel{\rlap{\lower4pt\hbox{\hskip1pt$\sim$}} \raise1pt\hbox{$<$}}} % less than or approx. symbol

\begin{document}

\title{Supercritical Dirac resonance parameters from extrapolated analytic continuation methods}
\begin{abstract}
The analytic continuation methods of complex scaling (CS), smooth exterior scaling (SES), and  complex absorbing potential (CAP) are investigated for the supercritical quasimolecular ground state in the U$^{92+}$-Cf$^{98+}$ system at an internuclear separation of $R=20$ fm. Pad\'e approximants to the complex-energy trajectories are used to perform an extrapolation of the resonance energies, which, thus, become independent of the respective stabilization
parameter. Within the monopole approximation to the two-center potential is demonstrated that the extrapolated results from SES and CAP are consistent to a high degree of accuracy. Extrapolated CAP calculations are extended to include dipole and quadrupole terms of the potential for a large range of internuclear separations $R$. These terms cause a broadening of the widths at 
the $\promille$ level when the nuclei are almost in contact, and at the \% level
for $R$ values where the 1S$\sigma$ state enters the negative continuum.
\end{abstract}
\author{Edward Ackad and Marko Horbatsch}
\affiliation{Department of Physics and Astronomy, York University,\\ 4700 Keele St, Toronto, Ontario, Canada M3J 1P3}

\maketitle

\newpage
\section{Introduction}
The ability to produce super-strong electric fields in experiments, either by super-intense lasers or in collisions with highly ionized heavy atoms, opens the possibility to study nonlinear phenomena. It allows for detailed testing of the interface of relativistic quantum
mechanics and quantum electrodynamics by probing electron dynamics at an energy scale where phenomena such as electron-positron pair creation may be detected against the background of more conventional atomic physics processes such as ionization and positron production by time-dependent fields \cite{pra10324}.

An electron in a Coulomb potential represents a problem that has been studied in both relativistic and non-relativistic physics. As the potential depth increases with nuclear charge $Z$, the ground state's energy decreases as $Z^2$. Since the spectrum of the Dirac equation for a Coulomb potential is not bounded from below (due to the negative-energy continuum $E<-$m$_{\mathrm{e}}$c$^2$), a potential of sufficient strength can yield resonance states when the ground-state energy $E_{\mathrm{1S}}<-$m$_{\mathrm{e}}$c$^2$. This causes the QED vacuum to become unstable and decay by pair-creation to a charged vacuum state \cite{rafelski}. The potential is called super-critical in this case. 

For a Coulomb potential, a single nucleus would need to be much more massive than any observed element ($Z \gsim 169$ for finite-size nuclei), but when two large nuclei get close enough, the combined two-center potential can become supercritical \cite{greiner,rafelski}. As the two nuclei get closer the energy of the quasi-molecular ground state (1S$\sigma$) decreases and so the resonance energy is embedded deeper into the negative-energy continuum of the Dirac spectrum. These negative-energy electron states can be reinterpreted using CPT symmetry as positron states of positive energy. Therefore, by studying these supercritical resonances, the dynamics of systems at the energy scale of particle creation can be explored.

It is possible to build a matrix representation of the hermitian two-center Hamiltonian for a supercritical potential in order to measure the resonance parameters from the Lorentzian distributions in the energy spectrum or from the density of states \cite{ackadcs}. More accurate results were obtained from analytic continuation methods which yield discrete resonance states with the following properties: {\it (i)} bounded, square-integrable eigenfunctions,i.e., elements of the Hilbert space of the original Hamiltonian; {\it (ii)} complex eigenenergies, $E_R$, whose real and imaginary parts agree with the center and width of the Lorentzian shape in the spectrum of the original Hamiltonian. Note that the imaginary part is positive, {\it i.e.}, $E_R=E_{\mathrm{res}}+i\Gamma/2$ for Dirac states with $E_{\mathrm{res}}<-$m$_{\mathrm{e}}$c$^2$ \cite{ackadcs}. 

Analytic continuation makes use of some parameter $\zeta$ to turn the original Hamiltonian into a non-hermitian operator. This raises two problems: {\it (i)} one has to optimize $\zeta$ to find the best approximation to the complex resonance eigenenergy; {\it (ii)} one worries about the effect of the unphysical $\zeta$ parameter on the final result. The optimization in step {\it (i)} is performed by stabilizing some measure that depends on the eigenenergy as a function of $\zeta$
(such as, e.g., the magnitude $|E_R(\zeta)|$). 

Recent work for the method of adding a complex absorbing potential (CAP) has demonstrated how to obtain more accurate results by extrapolating the complex $E_R(\zeta)$-trajectory to $\zeta=0$ \cite{lebrevePade}. The present work extends this idea to the relativistic Dirac equation using the complementary methods of smooth exterior scaling (SES) and CAP. The stability and parameter independence of the results is demonstrated. The extrapolation technique enables us to perform an extension of the calculations beyond the monopole approximation to the two-center potential. Results are given for the three coupled channels, $\kappa=-1,1,-2$, for a large range of internuclear separation, i.e., the resonance width broadening due to dipole and quadrupole interactions is calculated.

\section{Theory}
\subsection{Complex absorbing potential method}
The method of adding a complex absorbing potential (CAP) to the Hamiltonian has been used extensively in atomic and molecular physics \cite{ingrCAP,santraSO,santragf,sahoo}, and was put on firm mathematical grounds by Riss and Meyer \cite{RissCAP}. It was recently extended to the relativistic Dirac equation for Stark resonances \cite{CSDirac}, and for supercritical resonances \cite{ackadcs}. It works by extending the physical Hamiltonian by an imaginary potential which makes the Hamiltonian non-hermitian. In the case of the Dirac Hamiltonian a CAP is added as a scalar giving,
\begin{equation}
\hat{H}_{\mathrm{CAP}} = \hat{H}-i\eta \hat{\beta}{W}(r),
\end{equation}
where $\eta$ is a small non-negative parameter determining the strength of the CAP and $\hat{\beta}$ is the standard Dirac matrix \cite{greiner}. The function ${W}(r)$ determines the shape of the potential and is tailored for the specific problem. Currently, the most common use of a CAP is as a stabilization method: one solves the system on an equally spaced mesh of values of $\eta$ and takes the minimum of $\left|\eta \frac{dE_{R}}{d\eta}\right|_{\eta=\eta_{\mathrm{opt}}}$ as the closest approximation to the true resonance parameters \cite{RissCAP}.

In the present work we chose 
\begin{equation}\label{cap}
W(r)=\Theta(r-r_{\mathrm{c}}) \left(r-r_{\mathrm{c}}\right)^2
\end{equation}
for the CAP, where $\Theta$ is the Heaviside function. The CAP parameter $r_{\mathrm{c}}$ allows for the turn-on of the potential outside of the ``bound'' part of the wavefunction \cite{santra}.

\subsection{Smooth exterior scaling method}
Smooth exterior scaling (SES) is an extension of the complex scaling (CS) method whose mathematical justification was developed, e.g., by W.P. Reinhardt \cite{reinhardtcs}, and Moiseyev \cite{moiseyevrep}. CS introduces an analytical continuation of the Hamiltonian by scaling the reaction coordinate, $r$, by $r \rightarrow r e^{i \theta}$, and has been used extensively in atomic and molecular physics \cite{eHbreakupviaCS,RoyalCSmolecule,RoaoCSstark,CR4}. CS was recently extended to the relativistic Dirac equation for supercritical resonances \cite{ackadcs} and in 3D for a Coulomb potential with other short range potentials \cite{3drelcs}. SES relies on the same justification as CS, but uses a general path in the complex plane that is continuous;  when using non-continuous paths one refers to the method as exterior complex scaling (ECS) \cite{moiseyevrep}. A simple path is obtained by rotating the reaction coordinate into the complex plane about some finite position, $r_{\mathrm{s}}$, instead of the origin. The transformation then has the form,
\begin{equation}\label{ses}
r \rightarrow \left\{
\begin{array}{ccc}
r & & \mathrm{for}\;\; r<r_{\mathrm{s}} \\
\left(r-r_{\mathrm{s}}\right)e^{i\theta} + r_{\mathrm{s}} & & \mathrm{for}\;\; r_{\mathrm{s}} \leq r.
\end{array}\right.
\end{equation}
It offers the advantage of turning on the scaling at a distance $r_{\mathrm{s}}$ which can be chosen appropriately for a given potential shape. In analogy to the $r_{\mathrm{c}}$ parameter of the CAP method it is natural to choose $r_{\mathrm{s}}$ such that the ``bound'' part of the resonance state is not affected directly by the complex scaling.
 
This additional freedom introduces some complications not found in CS: unlike CS, SES does not always have a minimum in the $\left| \frac{dE_R}{d\theta} \right|$ curve as a function of $\theta$, which makes it difficult to determine a stabilized $\theta$ value, $\theta_{\mathrm{opt}}$, for $E_R(\theta)$. An approximation can always be made by finding the cusp of the trajectory in $\{E_{\mathrm{res}}, \Gamma\}$ space \cite{Doolencusp}, although this does not allow for a very precise determination of $\theta_{\mathrm{opt}}$ compared with a minimum in $\left| \frac{dE_{R}}{d\theta}\right|$. Alternatively, the parameters can be determined from either the $\frac{dE_{\mathrm{res}}}{d\theta}$ or $\frac{d\Gamma}{d\theta}$ curves, since in practice it is usually found that at least one of them will have a minimum, with the best results for each parameter coming from its own derivative minimum \cite{moiseyevcs}. Optimal results are obtained when the three derivatives have the same value of $\theta_{\mathrm{opt}}$ yielding the same values for  $E_{\mathrm{res}}$ and $\Gamma$(1S$\sigma$). 

Although CAP and SES appear as separate methods, they are not independent. It has been shown how a CAP can be transformed into CS \cite{santrawhycap} and that SES is related to CAP \cite{SES2cap}. It has also been shown that a transformative CAP (TCAP) gives the same Hamiltonian as SES with the unscaled potential and a (small) correction term \cite{RissSESCAP}.

\subsection{Pad\'e approximant and extrapolation}\label{padeexp}
The goal of the analytic continuation methods is to make the resonance wavefunction, $\psi_{\mathrm{res}}$, a bounded function by choosing an analytic continuation parameter $\zeta > \zeta_{\mathrm{crit}}$. Although $\psi_{\mathrm{res}}$ is an eigenfunction of the physical, hermitian, Hamiltonian, it is not in the Hilbert space since it is exponentially divergent. For a sufficiently large critical value of $\zeta=\zeta_{\mathrm{crit}}$ (called $\theta_{\mathrm{crit}}$ for CS, and $\eta_{\mathrm{crit}}$ for CAP), $\psi_{\mathrm{res}}$ becomes a bounded function and is therefore in the Hilbert space of the physical Hamiltonian \cite{moiseyevrep}. Taking the $\lim_{\zeta\rightarrow 0} E_{R}(\zeta)$ always yields a real eigenvalue corresponding to $\hat{H}(\zeta=0)\Psi=E_R(\zeta=0)\Psi$ since $\hat{H}(\zeta=0)$ is hermitian when acting on bounded functions.

The authors of Ref.~\cite{lebrevePade} proposed the following: instead of taking directly the $\zeta=0$ limit, an extrapolation of a part of the trajectory, $E_R(\zeta)$, namely for $\zeta > \zeta_{\mathrm{crit}}$, is used to obtain the complex-valued $E_{\mathrm{Pad\acute{e}}}(\zeta=0)$. The points used for the extrapolation are computed eigenvalues restricted to a region where  $\psi_{\mathrm{res}}$ remains in the Hilbert space
(as represented by the finite basis). For the extrapolation the Pad\'e approximant is given by,
\begin{equation}\label{pade}
E_{\mathrm{Pad\acute{e}}}(\zeta)= \frac{\sum_{i=0}^{N_1}p_i \zeta^i}{1+\sum_{j=1}^{N_1+1}q_j \zeta^j}
\end{equation}
where $p_i$ and $q_j$ are complex coefficients, and $N_p=2(N_1+1)$ is the number of points used in the approximant \cite{SchlessingerPade}. The extrapolated value of $E_{\mathrm{Pad\acute{e}}}(\zeta=0)$ is given simply by $p_0$, and follows from a given set of $\zeta$-trajectory points,
\begin{equation}\label{epoints}
\epsilon_i=\left\{ \varepsilon_j=E_R(\zeta_{i}+j\Delta\zeta),\;j=1...N_{\mathrm{p}} \right\},
\end{equation}
where $\zeta_{i} \geq \zeta_{\mathrm{crit}}$. For practical considerations about the added complications of the method (choice of $\epsilon_i$, order of the extrapolation) we refer the reader to section \ref{padeexpl}.

\section{Results}\label{resultssec}
To examine the smooth exterior scaling (SES) method and compare it to the simpler complex scaling (CS) method, and to the method of a complex absorbing potential (CAP), we look at the supercritical system of a single electron exposed to the field of a uranium ($A=238$) nucleus separated by $20 \ \mathrm{fm}$ from a californium ($A=251$) nucleus. The same system was explored in Ref.~\cite{ackadcs} using the CS and CAP methods and in Ref.~\cite{pra10324} using phase-shift analysis and numerical integration. The nuclei are approximated as displaced homogeneously charged spheres with a separation of $R$ between the charge centers. In the center-of-mass frame the monopole potential for each of the two nuclei with respective charge $Z_i$, radius $R_{\rm n}^{(i)}$ and center-of-mass displacement $R_{\mathrm{CM}}$ is given (in units of $\hbar =c =m_{\mathrm{e}}=1,Z=Z_i,R_{\mathrm{n}}=R_{\mathrm{n}}^{(i)} $) by 
\begin{equation}\label{pot} V(r)= \left\{ \begin{array}{cc}
-\frac{Z\alpha}{r} & \mathrm{for}\;r>r_+ \\

-\frac{Z\alpha}{R_{\mathrm{n}}^3 R_{\mathrm{CM}}}\left[ \frac{(R_{\mathrm{CM}}-R_{\mathrm{n}})^3(R_{\mathrm{CM}}+3R_{\mathrm{n}})}{16r}-\frac{r_+^2(R_{\mathrm{CM}}-2R_{\mathrm{n}})}{4}+\frac{3(R_{\mathrm{CM}}^2-R_{\mathrm{n}}^2)r}{8} - \frac{R_{\mathrm{CM}}r^2}{8} +\frac{r^3}{16}
\right] & \mathrm{for}\; r_- <r<r_+ \\

-\frac{Z\alpha}{R_{\mathrm{CM}}} & \mathrm{for}\;r <r_- 
\end{array} \right.
\end{equation}
where $r_{\pm} = R_{\mathrm{CM}}\pm R_{\mathrm{n}}$ \cite{2ctrCMpot}.
The expression is obtained from the potential for a homogeneously charged
sphere displaced by $R_{\mathrm{CM}}$ along the $z$-axis and then expanded in Legendre
polynomials.

To solve the Dirac equation for the potential given by Eq.~(\ref{pot}), a matrix representation is constructed using the mapped Fourier grid method \cite{me1}. The radial coordinate is mapped to a new variable $\phi$ by,
\begin{equation}\label{mapping}
r(\phi)=\frac{s\phi-4000\arctan{\frac{s\phi}{4000}}}{\left(\pi -\phi\right)^2},
\end{equation}
were $s$ is a scaling parameter which allows for the tailoring of the $N$ mesh points. The transformation maps $r \in [0,\infty)$ to $\phi \in [0,\pi)$ which allows for a more efficient coverage of the relevant phase space.

\subsection{Pad\'e extrapolation of SES trajectories}
\begin{figure}[!ht]
\includegraphics[angle=270,scale=0.5]{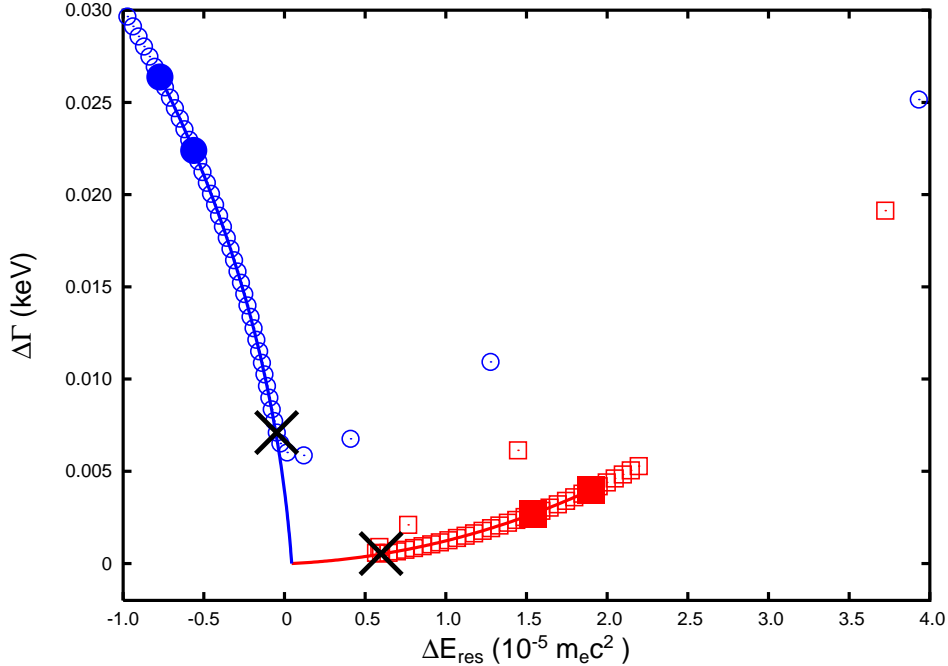}
\caption{\label{trajs} (Color Online) The $\Gamma(1{\mathrm S}\sigma)$ resonance $\theta$-trajectory from $N=250$, $r_{\mathrm{s}}=2$ (red squares) and $r_{\mathrm{s}}=3$ (blue circles) displayed as the difference from a reference calculation (cf. text) for the U-Cf system at $R=20$ fm in monopole approximation. The $\times$'s mark the resonance energy obtained by minimizing $\left|\frac{dE_R}{d\theta}\right|$. The $N_{\mathrm{p}}=8$ data points used for the extrapolations are bracketed by the solid symbols, and a spacing of  $\Delta\theta=0.01$ was used.}
\end{figure}

In SES (as in the other methods) the energy $E_R(\theta)$ forms a trajectory which depends on the calculation parameters ($N,s,r_{\mathrm{s}}$). Figure~\ref{trajs} shows the $\theta$-trajectory for $N=250$, $s=400$, $\Delta\theta=0.01$ and $r_{\mathrm{s}}=2$ and $r_{\mathrm{s}}=3$ respectively as a sequence of squares and circles  respectively.
The values are displayed as the deviation from a reference value given by $E_{\mathrm{res}}=-1.757930695(8)$m$_{\mathrm{e}}$c$^2$ and $\Gamma=4.084798(12)$keV, which was obtained by a large-size CAP calculation using $N=3000$, $s=400$ and $r_{\mathrm{c}}=2$. The uncertainties were determined from calculations with different basis parameters, $s$.

In Fig.~\ref{trajs}, the $E_{R}(\theta)$ results for small $\theta$ values start in the upper right hand and abruptly change direction after a few points. This happens in the vicinity of the reference value given by the point (0,0). As $\theta$ increases beyond the cusp in the trajectory, $E_R(\theta)$ changes more slowly, and the $\theta$-trajectory points become denser. Due to the finite number of collocation points $N$, the actual value of $\theta_{\mathrm{crit}}$ (the minimum $\theta$ value such that $\psi_{\mathrm{res}}$ is bounded and properly represented within the finite basis) is higher than theoretically predicted \cite{moiseyevrep}, and is close to the cusp-like behavior in the $\theta$-trajectory. 

Optimal, directly calculated, approximate values for the resonance energy are chosen by finding the minimum of $\left| \frac{dE_{R}}{d\theta}\right|_{\theta=\theta_{\mathrm{opt}}}$ \cite{moiseyevrep,reinhardtcs}, and are displayed as black crosses for both choices of $r_{\rm s}$. The solid lines represent $N_{\mathrm{p}}=8$ Pad\'e extrapolation curves (cf. Eq.~(\ref{pade})). The values of $\varepsilon_1$ and $\varepsilon_{N_{\mathrm{p}}}$ from Eq.~(\ref{epoints}), i.e., the bracketing points used to determine the Pad\'e approximant, are indicated by filled symbols. The values of $\theta_1=0.28$ for the $r_{\mathrm{s}}=2$ curve and $\theta_1=0.36$ for the $r_{\mathrm{s}}=3$ curve were chosen, because their $E_{\mathrm{Pad\acute{e}}}(\zeta=0)$ points were found to be closest to each other. Using other starting values for the Pad\'e curves would result in indistinguishable extrapolation curves on the scale of Fig.~\ref{trajs} (as long as the starting point would not be too close to the cusp in the $\theta$-trajectory). From the Pad\'e curves one obtains the best approximation to the resonance parameters by selecting the end points ($E_{\mathrm{Pad\acute{e}}}(\zeta=0)$), located close to the the origin (0,0), i.e., close to the $N=3000$ reference calculation result.

Although very different trajectories from those shown in Fig.~\ref{trajs} were obtained for different $r_{\mathrm{s}}$ values, the Pad\'e curves still extrapolated to nearly the same $\zeta=0$ resonance energy, as long as $r_{\mathrm{s}}$ was chosen outside the ``bound'' part of the wavefunction and not too large, i.e. $1.5 \lsim r_{\mathrm{s}}\lsim 6$. Larger values of $r_{\rm s}$ pose difficulties for the finite-$N$ computation as it cannot span the large-$r$ region properly. 

Simple complex scaling (CS) is equivalent to SES with $r_{\mathrm{s}}=0$ and, thus, falls outside of the range of acceptable $r_{\mathrm{s}}$ values. It yields results with a much larger deviation from the reference result, even when Pad\'e extrapolation is applied.

\begin{figure}[!ht]
\includegraphics[angle=270,scale=0.5]{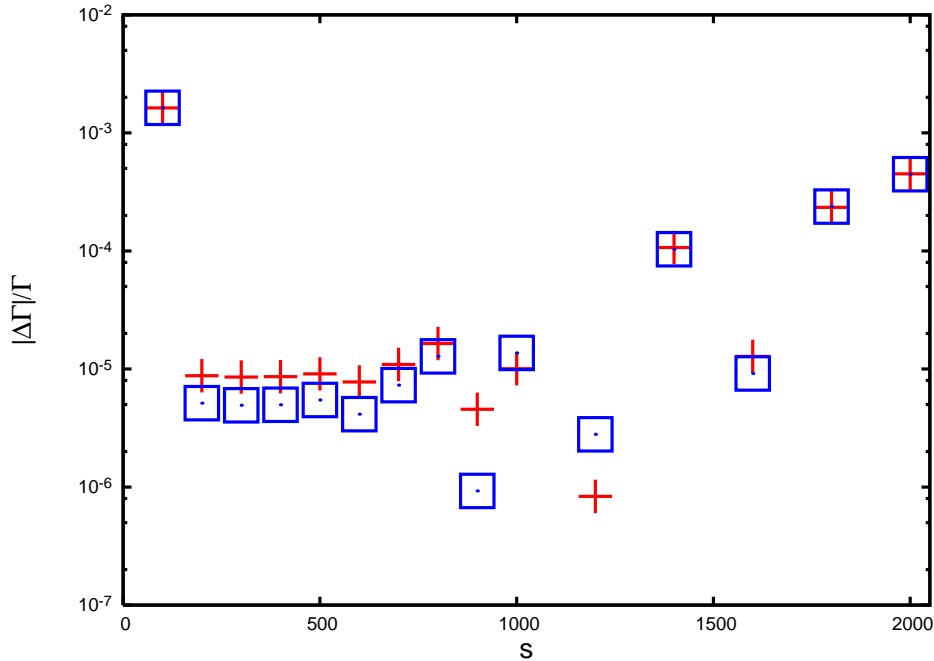}
\caption{\label{NS} (Color Online) Magnitude of relative error for Pad\'e extrapolated widths $\Delta \Gamma$(1S$\sigma$) for $N=500$ as a function of the scaling parameter, $s$, for the U-Cf system at $R=20$ fm in monopole approximation. Data shown are obtained using two reference calculations from the stabilized CAP method with $N=3000$ and $r_{\mathrm{c}}=2$; red plus symbols: $s=400$; blue squares: $s=750$ (cf. text). }
\end{figure}

The scaling parameter, $s$, introduced by the mapping of the radial coordinate in the mapped Fourier grid method in Eq.~(\ref{mapping}), plays a dominant role as far as numerical accuracy is concerned. In Fig.~\ref{NS} the magnitude of the relative error in the width $\Gamma$(1S$\sigma$) is shown as a function of $s$ for SES calculations with $N=500$, while using two reference CAP calculations. The latter were obtained from the stabilization method with a large basis ($N=3000$) and yielded values of $E_{\mathrm{res}}=-1.757930695$, $\Gamma=4.084798$ for scaling parameter $s=400$, and $E_{\mathrm{res}}=-1.757930702$, $\Gamma=4.084808$ for $s=750$ respectively.
Each SES data point was obtained by finding the closest intersection of the $r_{\mathrm{s}}=2,3,4$ Pad\'e curves from all starting $\theta$'s (for $\theta > \theta_{\mathrm{crit}}$), and taking the average of these three best values. The use of different $r_{\mathrm{s}}$ values within the acceptable range would cause changes too small to see on this plot. The resonance position, $E_{\mathrm{res}}$, is much more stable with respect to $s$ and therefore not shown.

The technique of combining information from different $r_{\mathrm{s}}$ calculations in order to select the ideal subset of $\theta$ points helps with the following problem. For each calculation with fixed value of $r_{\mathrm{s}}$ we looked at extrapolated values for the resonance width as a function of the start value $\theta_{i}$, and observed the deviation from the average value. It was found that numerical noise present in the data was minimal for some range of $\theta_{i}$ values (different for each calculation). This noise was investigated as a function of extrapolation order $N_{\mathrm{p}}$. Combining information from calculations with different values of $r_{\mathrm{s}}$ allowed us to eliminate two sources of random uncertainties associated with finite-precision input to a sensitive extrapolation calculation. 

The results demonstrate that the extrapolated ($\theta=0$) SES $N=500$ calculations exceed the precision of the stabilized (non-extrapolated) $N=3000$ CAP calculations. Table \ref{sestable} gives the resonance parameter results ($E_{\mathrm{res}},\Gamma$) for two different extrapolations, namely $N_{\mathrm{p}}=4,8$ for basis size $N=500$, and mapping parameter $300 \le s \le 700$, which corresponds to a subset of the data shown in Fig.~\ref{NS}. The standard deviation of the mean, 
\begin{equation}\label{std}
\sigma=\sum_i^{N_i}{\frac{\sqrt{(x_i-\bar{x})^2}}{N_i}},
\end{equation}
where $\bar{x}$ is the average $x$-value, are given separately for resonance position and width. The results are encouraging since extrapolation provides a  marked improvement over the stabilized CS method discussed in Ref.~\cite{ackadcs}.  
\begin{table}[htbp]
	\centering
\begin{tabular}{|c|c|c|c|c|}
\hline 
 $N_{\mathrm{p}}$ & $ E_{\mathrm{res}}$& $\sigma({E_{\mathrm{res}}})$ & $\Gamma$ (keV) & $\sigma({\Gamma})$ \\
\hline\hline
8 &-1.75793073 & 1.7$\times10^{-7}$ &	4.0848085 &	1.4$\times10^{-6}$ \\
\hline
4 & -1.75793072 & 1.6$\times10^{-7}$ & 4.0848031 &	1.4$\times10^{-6}$ \\
\hline
\end{tabular}
	\caption{\label{sestable}Averaged resonance position and width from the extrapolated SES method with basis size $N=500$ using extrapolation orders $N_{\mathrm{p}}=8$ and $N_{\mathrm{p}}=4$. The values were averaged over the stable range $300\le s\le 700$. The standard deviation of the mean is given by $\sigma$ separately for position and width.}
\end{table}

\subsection{Pad\'e extrapolation of CAP trajectories}\label{capcs}
We have shown that a quadratic complex absorbing potential (cf. Eq.~\ref{cap}) gives better results than simple complex scaling (CS) when used as a stabilization method \cite{ackadcs}. We find that the performance of SES represents an improvement over that of CS and it is competitive with CAP. 

\begin{figure}[!ht]
\includegraphics[angle=270,scale=0.5]{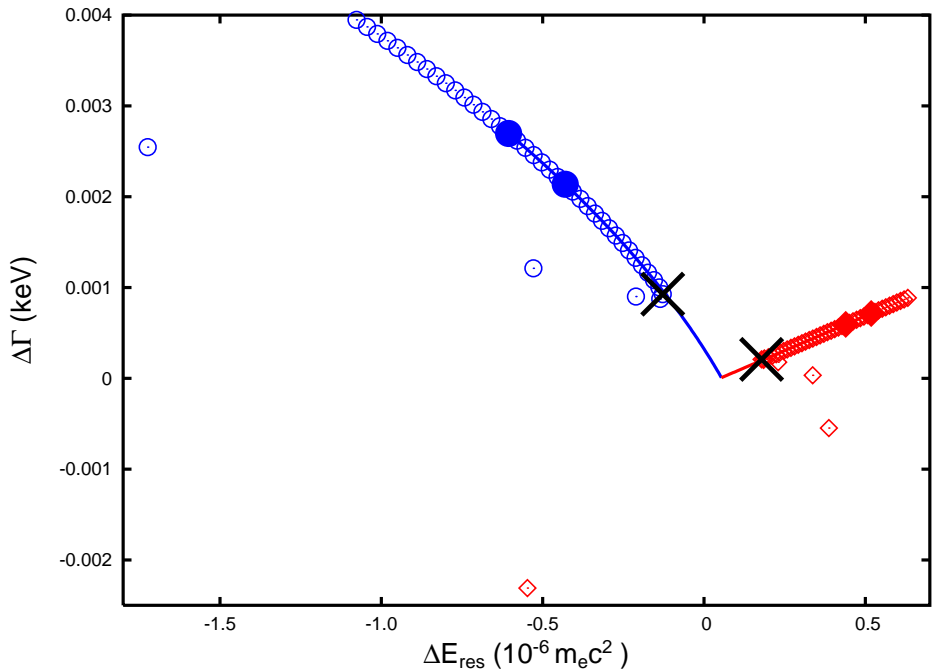}
\caption{ \label{trajs2}(Color Online) SES $\theta$-trajectory with $N=500$, $s=600$, $r_{\mathrm{s}}=3$ (blue circles), and CAP $\eta$-trajectory for $N=500$, $s=600$, $r_{\mathrm{c}}=3$ (red diamonds) displayed as the difference from a reference calculation for the 1S$\sigma$ state in the U-Cf system at $R=20$ fm in monopole approximation. The $\times$'s mark the approximate values obtained from the stabilization method. The reference value was obtained by Pad\'e extrapolation of CAP calculations with $N=3000$, $N_{\mathrm{p}}=4$ for $r_{\mathrm{c}}=2,3,4$ yielding $E_{\mathrm{res}}=-1.7579307062(5)$m$_{\mathrm{e}}$c$^2$ and $\Gamma=4.08481303(8)$keV.}
\end{figure}

Even for the same calculation parameters ($N,s,r_{\mathrm{c}}$), the $\eta$-trajectory for CAP is found to be different from the $\theta$ trajectory in SES. Figure~\ref{trajs2} shows a trajectory pair for identical calculation parameters, namely $N=500$, $s=600$, and $r_{\mathrm{s}}=r_{\mathrm{c}}=3$ respectively.
Even though the trajectories are rather different, the Pad\'e curves extrapolate to rather close values ($E_{\mathrm{Pad\acute{e}}}(\eta=0)$) as ($E_{\mathrm{Pad\acute{e}}}(\theta=0)$) from the SES method, i.e. the extrapolated resonance parameters for both methods are very similar. When looking at plots of the $N=500$ $\eta=0$ extrapolation results for CAP as a function of $s$, as was done in Fig.~\ref{NS} for SES, we observe almost identical results. In table~\ref{captable}, the average and standard deviation of the mean, $\sigma$ (cf. Eq.~\ref{std}), for the CAP results for $N=500$, $300 \le s \le 700$ are given for two different extrapolation sizes, $N_{\mathrm{p}}=4,8$. 

\begin{table}[htbp]
	\centering
\begin{tabular}{|c|c|c|c|c|}
\hline 
$N_{\mathrm{p}}$ & $E_{\mathrm{res}}$& $\sigma_{E_{\mathrm{res}}}$ & $\Gamma$ (keV) & $\sigma_{\Gamma}$ \\
\hline\hline
8 &-1.75793072 & 1.6$\times10^{-7}$ &	4.0848141&	3.1$\times10^{-6}$ \\
\hline
4 & -1.75793072 & 1.7$\times10^{-7}$ & 4.0848122 &	5.4$\times10^{-6}$ \\
\hline
\end{tabular}
	\caption{\label{captable}Same as in Table~\ref{sestable}, but for the extrapolated CAP method with basis size $N=500$.}
\end{table}

\subsection{Comparison of CAP and SES results}\label{padeexpl}
Both methods (CAP and SES) show similar behavior with respect to different aspects of the calculation. Both are relatively insensitive to the number of Pad\'e points, $N_{\mathrm{p}}$, used in the extrapolation. It was found that the results were unchanged, to the relevant precision, for $4\leq N_{\mathrm{p}} \leq 12$. The effects of changing the distance between the analytic continuation parameter ($\Delta\theta$ for SES and $\Delta\eta$ for CAP) were similarly found to be small compared with $s$. 

It was found that one could choose optimal $\zeta_i$ start values for the Pad\'e approximation by minimizing the deviation between $E_R(\zeta=0)$ from different $r_{\mathrm{s}}$ or $r_{\mathrm{c}}$ calculations. This is simple to implement, and works well for different values of $N_{\mathrm{p}}$. In this way one obtains results that are independent of the starting point for the rotation ($r_{\mathrm{s}}$ in SES) or the imaginary potential ($r_{\mathrm{c}}$ in CAP). The results are therefore very stable with respect to the analytic continuation parameters and depend only on the parameters from the mapped Fourier grid method. As shown in Fig.~\ref{NS} for SES (a very similar graph was obtained for CAP), there is a range of $s$ for which results are stable. For larger basis size $N$ the stable $s$-range increases making a judicious choice of $s$ less important. 

We have averaged the results from both the SES and CAP methods over the stable $s$ region, for a basis size of $N=500$ in tables \ref{sestable} \& \ref{captable}. The standard deviation for the width $\sigma(\Gamma)$ within either method is below $10^{-5}$, while the results differ at this level. It is, therefore, of interest to determine the reliability of the error estimate which is based upon basis parameter variations using larger-$N$ extrapolated calculations. Comparison of the width results with such an estimate based upon $N=3000$ CAP and SES calculations indicates that SES and CAP converge to the same value, closest to the CAP value given in table \ref{captable}.

Concerning the most appropriate order for the extrapolations it is worth noting that the CAP Pad\'e trajectories are rather straight in comparison with the ones for SES, and $N_{\mathrm{p}}=4$ might be more appropriate in this case. Nevertheless, we find no systematic improvement when going to the lower-order approximation (which might be deemed more stable with respect to numerical noise in the trajectory points).

\section{Coupled-channel calculations}
The ability to compute the resonance parameters to high precision with moderate basis size (e.g., $N=500$) allows us to explore the effects of higher multipoles which are present in the two-center interaction. We are not aware of prior investigations of such coupled-channel resonance calculations of supercritical Dirac states. While the effect on the resonance position is expected to be small, the sensitivity of the width to computational details (cf. the different results discussed in \cite{ackadcs}) indicates that some broadening of the resonance may occur.

To account for such two-center potential effects the wavefunction is expanded using spinor spherical harmonics, $\chi_{\kappa,\mu}$, 
\begin{equation}
\Psi_{\mu}(r,\theta,\phi)=\sum_{\kappa}{ \left(
\begin{array}{c}
G_{\kappa}(r)\chi_{\kappa,\mu}(\theta,\phi) \\
iF_{\kappa}(r)\chi_{-\kappa,\mu}(\theta,\phi)
\end{array}
\right)} \quad ,
\end{equation}
which are labeled by the relativistic angular quantum number $\kappa$ (analogous to $l$ in non-relativistic quantum mechanics) and the magnetic quantum number $\mu$ \cite{greiner}. The Dirac equation for the scaled radial functions, $f(r)=rF(r)$ and $g(r)=rG(r)$, then becomes ($\hbar=c=1$),
\begin{eqnarray}
\label{syseqn}
\frac{df_{\kappa}}{dr} - \frac{\kappa }{r}f_{\kappa} & = & - \left(E -1 \right)g_{\kappa} + \sum_{\bar{\kappa}=\pm1}^{\pm\infty}{\langle \chi_{\kappa,\mu} \left| V(r,R) \right| \chi_{\bar{\kappa},\mu} \rangle }g_{\bar{\kappa}} \quad , \\
\label{syseqn2}
\frac{dg_{\kappa}}{dr} + \frac{\kappa}{r}g_{\kappa} & = & \left(E + 1 \right) f_{\kappa} - \sum_{\bar{\kappa}=\pm1}^{\pm\infty}{\langle \chi_{-\kappa,\mu} \left| V(r,R) \right| \chi_{-\bar{\kappa},\mu} \rangle }f_{\bar{\kappa}} \quad ,
\end{eqnarray}
where $V(r,R)$ is the potential for two uniformly charged spheres displaced along the $z$-axis, which is expanded into Legendre polynomials according to $V(r,R)=\sum^{\infty}_{l=0}{V_l(r,R) P_l(\cos{\theta})}$ \cite{greiner}. The monopole term, $V_0$, is given explicitly in Eq.~(\ref{pot}), and for the present work we include the coupling terms required for the $\kappa=\pm1,-2$ channels (i.e., the $V_1(r,R)\langle \chi_{\pm\kappa,\mu} \left| P_1 \right| \chi_{\pm\bar{\kappa},\mu} \rangle$ dipole and $V_2(r,R)\langle \chi_{\pm\kappa,\mu} \left| P_2 \right| \chi_{\pm\bar{\kappa},\mu} \rangle$ quadrupole terms). The $\kappa=1,-2$ channels ($P_{1/2}$ and $P_{3/2}$ respectively) have the strongest coupling to the $\kappa=-1$ (S-states) and are therefore  expected to have the largest impact on the supercritical ground state.

\begin{table}[hthp]
	%\centering
\begin{tabular}{|c||r@{.}l|r@{.}l||r@{.}l|c||c|}
\hline
\multirow{2}{*}{$R$(fm)} & \multicolumn{4}{|c||}{Single-channel ($\kappa=-1$)} & \multicolumn{3}{|c||}{Three-channel ($\kappa=-1,1,2$)} & $\frac{|\Gamma_1-\Gamma_3|}{\Gamma_1}$ \\ 
\cline{2-8}
& \multicolumn{2}{|c|}{$E_{\mathrm{res}}$ (mc$^2$)}& \multicolumn{2}{|c||}{$\Gamma_1$ (keV)} & \multicolumn{2}{|c|}{$E_{\mathrm{res}}$ (mc$^2$)}& $\Gamma_3$ (keV) &  ($\times10^{-3}$)\\
\hline\hline
16 & -2&00635363(3) &  8&148233(4) & -2&00646180(3) & 8.150153(3) & 0.236\\
18 & -1&87487669(4) &  5&881605(1) & -1&87502343(4) & 5.883977(2) &	0.403 \\
20 & -1&75793073(4) &  4&0848148(4)& -1&75811259(4) & 4.087394(1) & 0.631\\
22 & -1&65393272(3) &  2&708987(1) & -1&65414448(3) & 2.711524(2) & 0.937\\
24 & -1&5612122(1)  &  1&6966694(3)& -1&56144826(1) & 1.698947(1) & 1.34\\
26 & -1&47820830(4) &  0&9877052(4)& -1&47846358(4) & 0.989571(1) & 1.89\\
28 & -1&40354329(5) &  0&5221621(4)& -1&40381341(4) & 0.523544(1) & 2.65\\
30 & -1&33603643(6) &  0&2420749(2)& -1&33631779(6) & 0.242976(1) & 3.72\\
32 & -1&27469286(5) & 0&0931429(1)& -1&27498253(5) &  0.093639(5) & 5.32\\
34 & -1&21868219(5) & 0&0271161(1)& -1&21897787(4) &  0.027322(2) & 7.59\\
36 & -1&16731491(6) & 0&0050374(1)& -1&16761472(7) &  0.005091(2)& 10.7\\
38 & -1&12001745(6) & 0&0004170(3)& -1&12031983(6) &  0.000439(4) & 52.5 \\
\hline
\end{tabular}
	\caption{\label{3captable}Averaged $1{\mathrm S}\sigma$ resonance position and width from the extrapolated CAP method with basis size $N=500$, $N_{\mathrm{p}}=4$, for single-channel (monopole) and three-channel calculations as a function of separation, $R$, in the U-Cf system. The values were averaged over the stable range $300\le s\le 700$ using $r_{\mathrm{c}}=2,3,4$. The value in parenthesis represents the standard deviation from the average and the last column gives the relative difference of the width from three- to one-channel.}
\end{table}

Table \ref{3captable} gives the $1{\mathrm S}{\sigma}$ resonance parameters for different internuclear separations for both the one- and three-channel CAP calculations. The CAP calculations were performed using a basis size of $N=500$ per channel with $300\le s\le 700$. The CAP parameters of $r_{\mathrm{c}}=2,3,4$ were used to obtain the most stable $\theta$ range for extrapolation, which was carried out with order $N_{\mathrm{p}}=4$. Values in parentheses indicate the standard deviation of the mean (cf. Eq.~\ref{std}). We note that the accuracy of the calculations - as indicated by the deviation - is much higher than what is required to measure the effect of the the P-state channels on the $1{\mathrm S}{\sigma}$ resonance. 

The final column contains the relative difference of the width between the three-channel results and the monopole approximation. The correction due to the dipole and quadrupole potentials is seen to increase with internuclear separation $R$. This trend can be expected, as the overall interaction becomes less spherically symmetric in this limit. The effect is small, however, as the strongest contribution towards the binding energy (and thus the supercriticality) is provided by the monopole part. The dipole interaction contributions to the resonance parameters are a result of the relatively small charge asymmetry in the U$^{92+}$-Cf$^{98+}$ potential. Quadrupole couplings to D states can be expected to generate more significant changes in $E_{\rm res}$ and $\Gamma$.

\section{Discussion}
We have extended the method of smooth exterior scaling (SES) to the relativistic Dirac equation. For both the addition of a complex absorbing potential (CAP) and SES, we have also shown that by extrapolation of the complex energy eigenvalue trajectories which are functions of the analytic continuation parameter, $\zeta$, to $\zeta=0$ using appropriate $\zeta$-trajectory values (cf. section~\ref{padeexp}), one obtains an accurate and robust estimation of the resonance parameters. These results use much smaller basis than would be required for similar precision using stabilization. While adding the extrapolation adds new parameters, we found minimizing the distance between extrapolated results for different calculation parameters was a simple and highly effective method of reducing the parameter space. The effects of the number of calculation points, $N_{\mathrm{p}}$, was found to be very small provided it was in the range $4 \leq N_{\mathrm{p}} \leq 12$.

The effects of higher order terms and the $P$-states were explored and found to have a increasing influence as the two-center system becomes less spherically symmetric which occurs at larger internuclear separations $R$. The resonances could acquire additional broadenings and shifts from dynamical effects not included in the quasi-static approximation.

\section{Acknowledgments}
This work was supported by NSERC Canada, and was carried out using the Shared Hierarchical Academic Research Computing Network (SHARCNET:www.sharcnet.ca). E. Ackad was supported by the Ontario Graduate Scholarship program.
\bibliography{SES_3cCAP}
\end{document}